\documentstyle[preprint,aps]{revtex}
\tightenlines
\newcommand{\lsim}{\mathrel{\mathop{\kern 0pt \rlap
  {\raise.2ex\hbox{$<$}}}
  \lower.9ex\hbox{\kern-.190em $\sim$}}}
\newcommand{\gsim}{\mathrel{\mathop{\kern 0pt \rlap
  {\raise.2ex\hbox{$>$}}}
  \lower.9ex\hbox{\kern-.190em $\sim$}}}

\newcommand{\beq}     {\begin{equation}}
\newcommand{\eeq}     {\end{equation}}
\newcommand{\es}      {\epsilon}
\newcommand{\ev}      {\equiv}
\newcommand{\M}       {{\mathcal M}}
\newcommand{\no}      {\nonumber}

\begin{document}
\draft
\preprint{
\vbox{\hbox{AS-ITP-2001-017}
      \hbox{hep-ph/0110119}}
}
\title{
A Phenomenological Study on Lepton Mass Matrix Textures
}

\author{
Chun Liu$^{\:a}$ and Jeonghyeon Song$^{\:b}$
}

\vspace{1.5cm}

\address{
$^a$Institute of Theoretical Physics, Chinese Academy of Sciences,\\
P.O. Box 2735, Beijing 100080, China\\
$^b$Korea Institute for Advanced Study, 207-43 Cheongryangri-dong,
Dongdaemun-gu,\\ 
Seoul 130-012, Korea
}

\maketitle
\thispagestyle{empty}
\setcounter{page}{1}
\begin{abstract}
  The three active light neutrinos are used to explain the neutrino 
oscillations.  The inherently bi-large mixing neutrino mass matrix and 
the Fritzsch type, bi-small mixing charged lepton mass matrix are assumed.  
By requiring the maximal $\nu_\mu-\nu_\tau$ mixing for the atmospheric 
neutrino problem and the mass-squared difference approperiate for the 
almost maximal mixing  
solution to the solar neutrino problem, the following quantities are 
predicted: the $\nu_e-\nu_\mu$ mixing, $V_{e3}$, CP violation in neutrino 
oscillations, and the effective electron-neutrino mass relevant to 
neutrinoless double beta decays.

\end{abstract}
\vspace{1.5cm}

\hspace{1.1cm}Keywords: lepton, neutrino oscillation, CP violation.

\pacs{PACS numbers: 14.60.Pq, 14.60.-z, 12.15.Ff.}

\newpage

Understanding the fermion mass pattern is a great challenge in 
elementary particle physics.  Lacking of a standard theory for the 
flavor physics, phenomenological ansatz might be very helpful \cite{fx}.  
In view of the recent observation about neutrino oscillations \cite{n}, 
this paper studies the lepton sector.  The masses of charged leptons 
have been known experimentally quite well \cite{pdg}.  They are expected 
to have a similar origin as quarks which have small mixings among three 
generations.

The small neutrino masses indicated by experiments can be naturally 
understood by the seesaw mechanism \cite{seesaw}.  However, the 
observations have shown increasing evidence that leptonic mixings are 
bi-maximal, or almost bi-maximal among the three generations.  Such a 
mixing scenario were then considered variously \cite{bim1,bim2,bim3}.  

This paper starts from the flavor eigenstates of both charged leptons 
and neutrinos.  We assume that the charged lepton mass matrix is of 
the Fritzsch type \cite{f}, namely, 
\beq
\label{1}
\M_l = \left(
\begin{array}{ccc}
0             & ae^{i\alpha} & 0           \\
ae^{-i\alpha} & 0            & be^{i\beta} \\
0             & be^{-i\beta} & c           \\
\end{array}\right)\,,
\eeq
where $c\gg b\gg a>0$ and $a<b^2/c$.  And the neutrino mass matrix is of 
the inherently bi-large mixing type \cite{bim1}, 
\beq
\label{2}
\M_\nu = \left(
\begin{array}{ccc}
\es & m_1 & m_2 \\
m_1 & \es & 0   \\
m_2 & 0   & \es \\
\end{array}\right)\,,
\eeq
where $m_1\sim m_2 \gg \es >0$.  Note that $m_1, ~m_2$ and $\es$ are 
always real in the above form of $\M_\nu$.  
These two matrices are of simplicity in the analysis, and the parameters 
in them are uniquely fixed.  Although Eq. (\ref{2}) will be speculated 
further in the end of this paper, we still have no definite principles 
for them.  Some more theoretical works for the bi-maximal leptonic 
mixing were considered in Ref. \cite{bim1,bim2,bim3}.  

The mass matrix Eq. (\ref{1}) gives
\beq
\label{3}
\begin{array}{ccl}
a&=&\displaystyle \left(\frac{m_e m_\mu m_\tau}{m_e-m_\mu+m_\tau}
\right)^{1/2}\,, \\[3mm] \no
b&=&\displaystyle \left(m_\mu m_\tau +m_\mu m_e-m_em_\tau-
\frac{m_e m_\mu m_\tau}{m_e-m_\mu+m_\tau}\right)^{1/2}\,, \\[3mm] \no
c&=&m_e-m_\mu+m_\tau\,.  
\end{array}
\eeq
Eq. (\ref{2}) gives neutrino masses, 
\beq
\label{4}
\begin{array}{ccl}
m_{\nu_1}&=&-\sqrt{m_1^2+m_2^2}+\es\,, \\ \no
m_{\nu_2}&=& \sqrt{m_1^2+m_2^2}+\es\,, \\ \no
m_{\nu_3}&=& \es\,.  
\end{array}
\eeq  

Charged leptons provide bi-small mixing among the three generations, 
whereas neutrinos provide bi-large mixing.  The diagonalization of 
$\M_l$ is made by the following unitary matrix \cite{fx2},
\beq
\label{5}
U_l = \left(
\begin{array}{ccc}
U_{11}^l             & U_{12}^l             & U_{13}^l \\
U_{21}^le^{-i\alpha} & U_{22}^le^{-i\alpha} & U_{23}^le^{-i\alpha}\\
U_{31}^le^{-i(\alpha+\beta)} & U_{32}^le^{-i(\alpha+\beta)} & 
U_{33}^le^{-i(\alpha+\beta)}  \\
\end{array}\right)\,,
\eeq
where
\beq
\label{6}
\begin{array}{ccl}
U_{11}^l&=&\displaystyle
        \left[1+\left(\frac{m_e}{a}\right)^2+\left(\frac{b}{a}
        \frac{m_e}{m_\tau-m_\mu}\right)^2\right]^{-1/2}\,, \\[3mm]\no
U_{22}^l&=&\displaystyle
        \left[1+\left(\frac{a}{m_\mu}\right)^2+\left(\frac{b}
        {m_\tau+m_e}\right)^2\right]^{-1/2}\,,             \\[3mm]\no
U_{33}^l&=&\displaystyle
        \left[1+\left(\frac{m_\mu-m_e}{b}\right)^2+
        \left(\frac{a}{b}\frac{m_\mu-m_e}{m_\tau}\right)^2
        \right]^{-1/2}\,,                                  \\[3mm]\no 
U_{12}^l&=&\displaystyle
        -\frac{a}{m_\mu}U_{22}^l\,, \\[3mm]\no
U_{13}^l&=&\displaystyle
        \frac{a}{b}\frac{m_\mu-m_e}{m_\tau}U_{33}^l\,,  \\[3mm]\no
U_{23}^l&=&\displaystyle
        \frac{m_\mu-m_e}{b}U_{33}^l\,, \\[3mm]\no
U_{21}^l&=&\displaystyle\frac{m_e}{a}U_{11}^l\,, \\[3mm]\no
U_{31}^l&=&\displaystyle
        -\frac{b}{a}\frac{m_e}{m_\tau-m_\mu}U_{11}^l\,, \\[3mm]\no
U_{32}^l&=&\displaystyle-\frac{b}{m_\mu+m_e}U_{22}^l\,.
\end{array}
\eeq 
$\M_\nu$ is diagonalized by 
\beq
\label{7}
U_\nu = \left(
\begin{array}{ccc}
\displaystyle\frac{1}{\sqrt{2}} &\displaystyle -\frac{1}{\sqrt{2}}
&0\\[3mm]
\displaystyle\frac{\sin\theta}{\sqrt{2}} 
&\displaystyle\frac{\sin\theta}{\sqrt{2}} &-\cos\theta\\[3mm]
\displaystyle
\frac{\cos\theta}{\sqrt{2}} &\displaystyle\frac{\cos\theta}{\sqrt{2}} 
&\sin\theta \\
\end{array}\right)\,,
\eeq
where $\sin\theta=\displaystyle\frac{m_1}{\sqrt{m_1^2+m_2^2}}$. 
Note that $U_\nu$ is independent on $\es$.  
The physical lepton mixing is given by 
\beq
\label{8}
V=U_l^{\dagger}U_\nu\,.
\eeq
It is the combination of the large mixing from $U_\nu$ and the small 
mixing from $U_l$ that gives the maximal mixing of $\nu_\mu-\nu_\tau$.  
In our scenario, $\cos\theta$ deviates from 
$\displaystyle\pm\frac{1}{\sqrt{2}}$ remarkably.  This is because the 
$(23)$ component of $V$ is mainly composed of $\cos\theta$ and 
$U_{23}^l\sim\sqrt{m_\mu/m_\tau}\sim 0.3$ which is not negligible.  
On the other hand, the matrix $U_\nu$ itself will give a maximal mixing 
in the $\nu_e-\nu_\mu$ oscillation, because the charged lepton 
contribution to $V_{12}$ is only about $\sqrt{m_e/m_\mu}\sim 0.01$.

Let us discuss the numerical results.  The quantity $\sqrt{m_1^2+m_2^2}$ is 
taken to be $0.05$ eV as indicated by the atmospheric neutrino problem.  By 
requiring the maximal $\nu_\mu-\nu_\tau$ mixing, we obtain
\beq
\label{9}
m_1\simeq 4.3\times 10^{-2}~{\rm eV,}~~~m_2\simeq 2.5\times 10^{-2}~
{\rm eV}\,.
\eeq
The solar neutrino problem is solved by the energy independent 
solution \cite{ei} which needs
\beq
\label{10}
|\es|\simeq 10^{-3}-10^{-4}~{\rm eV}~~{\rm or}~~
10^{-6}-10^{-8}~{\rm eV}\,.
\eeq
The $\nu_e-\nu_\mu$ mixing deviates from the maximal one slightly.  With 
the above results, we get 
\beq
\label{11}
\sin^2 2\theta_{e\mu}\simeq 0.99\,.
\eeq
The $\nu_e-\nu_\tau$ mixing is predicted as 
\beq
\label{12}
|V_{e3}|\simeq 0.049\,.
\eeq

The CP violation in the neutrino oscillations is determined by the 
rephasing-invariant parameter $J$ \cite{jw}, 
\beq
\label{13}
{\rm Im}(V_{i\lambda}V_{j\rho}V_{i\rho}^*V_{j\lambda}^*)=
J\sum_{k,\delta}\es_{ijk}\es_{\lambda\rho\delta}\,.
\eeq
In our case, Eqs. (\ref{5}$-$\ref{8}) give 
\beq
\begin{array}{ccl}
\label{14}
J&=&\displaystyle
    \frac{U^l_{12}}{2}(-U^{l2}_{11}+U^{l2}_{21}\sin^2\theta+U^{l2}_{31}
    \cos^2\theta+U^l_{21}U^l_{31}\sin 2\theta\cos\beta)\\[3mm]\no
  &&\times [U^l_{22}\sin\theta\sin\alpha
    +U^l_{32}\cos\theta\sin(\alpha+\beta)]\\[3mm]\no
  &&\displaystyle 
    -\frac{U^l_{11}}{2}(-U^{l2}_{12}+U^{l2}_{22}\sin^2\theta+U^{l2}_{32}
    \cos^2\theta+U^l_{22}U^l_{32}\sin 2\theta\cos\beta)\\[3mm]\no
  &&\times [U^l_{21}\sin\theta\sin\alpha
    +U^l_{31}\cos\theta\sin(\alpha+\beta)]\\[3mm]\no
 &\simeq&\displaystyle
    \frac{1}{4}\sqrt{\frac{m_e}{m_\mu}}\cos\theta\left\{\sin 2\theta
    \sin\alpha-2\sqrt{\frac{m_\mu}{m_\tau}}[\sin(\alpha+\beta)-2\sin^2
    \theta\sin\alpha\cos\beta ]\right\}\,.
\end{array}
\eeq 
Numerically, choosing $\alpha=\beta=\frac{\pi}{2}$, we can get 
$J\simeq 0.008$; choosing $\alpha=0$ and $\beta=\frac{\pi}{2}$, 
$J\simeq 0.004$.  

The neutrinoless double beta decay experiments will measure the effective 
electron-neutrino mass 
\beq
\label{15}
\langle m_{\nu_e}\rangle\ev|\sum_\lambda V_{e\lambda}^2m_{\nu_\lambda}| 
\eeq
which, by keeping $\es$ terms to the leading order, in our case is 
\beq
\begin{array}{ccl}
\label{16}
\langle m_{\nu_e}\rangle
 &=&2\sqrt{m_1^2+m_2^2}U^l_{11}[(U^l_{21}\sin\theta+U^l_{31}
    \cos\theta\cos\beta)^2+(U^l_{31}\cos\theta\sin\beta)^2]^{1/2}
    -\es U^{l2}_{11}\\[3mm]\no
 &\simeq&\displaystyle
 2\sqrt{m_1^2+m_2^2}\sqrt{\frac{m_e}{m_\mu}}\sin\theta-\es  \\[3mm] \no
 &\simeq&0.006~{\rm eV}\,.
\end{array}
\eeq

Experiments in the near future will check the reality of the lepton mass 
matrices studied in this paper.  In addition to SNO, Borexino and KamLAND 
will check the result of Eq. (\ref{11}) for the $\nu_e-\nu_\mu$ mixing 
\cite{v}.  The 
long baseline neutrino experiments \cite{beijing} and neutrino factories 
will measure $V_{e3}$ and CP violation in neutrino oscillations.  GENIUS 
is able to test the $\langle m_{\nu_e}\rangle$ given in Eq. (\ref{16}).  

Finally let us look at the underlying reasons of the neutrino mass matrix 
in Eq. (\ref{2}).  These Majorana masses are thought to be generated by 
the seesaw mechanism.  It is natural to assume that the Dirac neutrino 
mass matrix has similar form as that of charged leptons, 
\beq
\label{17}
{\cal \M_D} = \left(
\begin{array}{ccc}
0         & \tilde{a} & 0         \\
\tilde{a} & 0         & \tilde{b} \\
0         & \tilde{b} & \tilde{c} \\
\end{array}\right)\,, 
\eeq
where the possible phases are not considered, because $\M_\nu$  of 
Eq. (\ref{2}) is real and what we are looking at is magnitudes of 
righ-handed neutrino masses.  In this case, the texture of 
Eq. (\ref{2}) requires the following form of the right-handed neutrino 
mass matrix, 
\beq
\label{18}
\begin{array}{ccl}
{\cal \M_R} &=& \displaystyle\frac{1}{\es}\left(
\begin{array}{ccc}
\tilde{a}^2\cos^2\theta & -\tilde{a}\tilde{b}\sin\theta\cos\theta 
 & -\tilde{a}\cos\theta(\tilde{c}\sin\theta-\tilde{b}\cos\theta) \\
-\tilde{a}\tilde{b}\sin\theta\cos\theta & \tilde{b}^2\sin^2\theta 
 & \tilde{b}\sin\theta(\tilde{c}\sin\theta-\tilde{b}\cos\theta)  \\
-\tilde{a}\cos\theta(\tilde{c}\sin\theta-\tilde{b}\cos\theta) 
 & \tilde{b}\sin\theta(\tilde{c}\sin\theta-\tilde{b}\cos\theta) 
 & (\tilde{c}\sin\theta-\tilde{b}\cos\theta)^2 \\
\end{array}\right)\\[1cm] \no
            & &\displaystyle 
            +\frac{\tilde{a}}{\sqrt{m_1^2+m_2^2}}\left( 
\begin{array}{ccc}
0                   & \tilde{a}\sin\theta                     & 0 \\
\tilde{a}\sin\theta & 2\tilde{b}\cos\theta 
 & \tilde{c}\cos\theta+\tilde{b}\sin\theta \\
0                   & \tilde{c}\cos\theta+\tilde{b}\sin\theta & 0\\
\end{array}\right)\,.
\end{array}
\eeq
Note that in the above equation, the first matrix is the leading one.  But 
it is of rank one.  Only with the second matrix which is a perturbation to 
the first, is $\M_R$ nonsingular.  In the right-handed neutrino spectrum, 
there is a heavy one with mass around $(10^{15}-10^{16})$ GeV, and there 
are two relatively light neutrinos which are about two orders smaller than 
the first, if we take $\es\sim 10^{-4}$ eV.  
The form of $\M_R$ seems that some tuning is needed in order to 
keep the form of the texture assumed in Eq. (\ref{2}).  We wonder if there 
is a natural way to produce it, for instance from some flavor symmetry.  
\vspace{1.5cm}

\acknowledgments
We would like to thank Zhi-Zhong Xing for helpful discussions and comments.  
C.L. was supported in part by the National Natural Science Foundation of 
China with grant no. 10047005.


\begin{thebibliography}{99}

\bibitem{fx}
For a review, see H. Fritzsch and Z.-Z. Xing, Prog. Part. Nucl. Phys. 45 
(2000) 1.

\bibitem{n}
Super-Kamiokande Collaboration, Y. Fukuda {\it et al.}, 
Phys. Rev. Lett. 81 (1998) 1562;\\
Super-Kamiokande Collaboration, S. Fukuda {\it et al.}, 
Phys. Rev. Lett. 86 (2001) 5656;\\
SNO Collaboration, Q.R. Ahmad {\it et al.}, 
Phys. Rev. Lett. 87 (2001) 071301;\\
A. Habig, hep-ex/0106025, to appear in the Proceedings of ICRC 2001.

\bibitem{pdg}
Particle Data Group, D.E. Groom {\it et al.}, Eur. Phys. J. C15 (2000) 1.

\bibitem{seesaw}
M. Gell-Mann, P. Ramond and R. Slansky, in {\it Supergravity} 
(North Holland, 1979);
T. Yanagida, in {\it Proc. of the Workshop on Unified Theory and the 
Baryon Number of the Universe} (KEK, Tsukuba, 1979);
R. N. Mohapatra and G. Senjanovic, Phys. Rev. Lett. 44 (1980) 912.

\bibitem{bim1}
R. Barbieri, L.J. Hall, D. Smith, A. Strumia and N. Weiner, JHEP 9812 (1998)
017; For a review, see G. Altarelli and F. Feruglio, in Venice 1999, Neutrino
telescopes, p353.  

\bibitem{bim2}
W. Grimus and L. Lavoura, hep-ph/0110041;\\
E. Ma, D.P. Roy and S. Roy, hep-ph/0110146.  

\bibitem{bim3}
V. Barger, S. Pakvasa, T.J. Weiler and K. Whisnant, 
Phys. Lett. B 437 (1998) 107;
H. Fritzsch and Z.-Z. Xing, Phys. Lett. B 372 (1996) 265;
E. Torrente-Lujan, Phys. Lett. B 389 (1996) 557;
F. Vissani, hep-ph/9708483;
D.V. Ahluwalia, Mod. Phys. Lett. A 13 (1998) 2249;
M. Jezabek and Y. Sumino, Phys. Lett. B 440 (1998) 327;
R.N. Mohapatra and S. Nussinov, Phys. Lett. B 441 (1998) 299;
S. Davidson and S.F. King, Phys.Lett. B 445 (1998) 191;
A. Baltz, A.S. Goldhaber and M. Goldhaber, Phys. Rev. Lett. 81 (1998) 5730;
M. Fukugita, M. Tanimoto and T. Yanagida, Phys. Rev. D 57 (1998) 4429;
S. Davidson and S.F. King, Phys. Lett. B 445 (1998) 191;
C. Jarlskog {\it et al.}, Phys. Lett. B 449 (1999) 240; 
C. Giunti, Phys. Rev. D 59 (1999) 077301;
S.K. Kang and C.S. Kim, Phys. Rev. D 59 (1999) 091302;
H.B. Benaoum and S. Nasri, Phys. Rev. D 60 (1999) 113003;
Y.-L. Wu, Eur. Phys. J. C 10 (1999) 491;
H. Georgi and S. Glashow, Phys. Rev. D 61 (2000) 097301;
S.M. Barr and I. Dorsner, Nucl. Phys. B 585 (2000) 79;
C.S. Kim and J.D. Kim, Phys. Rev. D 61 (2000) 057301; 
M.C. Gonzales-Garcia, Y. Nir, A. Smirnov, C. Pena-Garay, 
Phys. Rev. D 63 (2001) 013007;
C.H. Albright and S.M. Barr, Phys. Rev. D 64 (2001) 073010;
M.-C. Chen and K.T. Mahanthappa, Phys. Rev. D 62 (2000) 113007;
Y. Koide and A. Ghosal, Phys. Rev. D 63 (2001) 037301;
Z.-Z. Xing, Phys. Rev. D 64 (2001) 093013;
D. Falcone, hep-ph/0106286;
K. Choi {\it et al.}, hep-ph/0107083;
W.J. Marciano, hep-ph/0108181;
M. Buchmuller and D. Wyler, hep-ph/0108216;
A. Aranda, C.D. Carone and P. Meade, hep-ph/0109120;
M. Lindner, T. Ohlsson and G. Seidl, hep-ph/0109264.

\bibitem{f}
H. Fritzsch, Phys. Lett. B 73 (1978) 317.

\bibitem{fx2}
For an example, see H. Fritzsch and Z.-Z. Xing, Nucl. Phys. B 556 (1999) 49. 

\bibitem{ei}
For recent studies, see 
S. Choubey, S. Goswami and D.P. Roy, hep-ph/0109017;\\
P. Greminelli, G. Signorelli and A. Strumia, JHEP 0105 (2001) 52.  

\bibitem{jw}
C. Jarlskog, Phys. Rev. Lett. 55 (1985) 1039;\\
D.-D. Wu, Phys. Rev. D 33 (1986) 860. 

\bibitem{v}
A. Strumia and F. Vissani, hep-ph/0109172.  

\bibitem{beijing}
For example, see H. Chen {\it et al.}, hep-ph/0104266.

\end{thebibliography}
\end{document}